\documentclass[11pt]{article}
\pdfoutput=1
\usepackage{amsmath}
\usepackage{amssymb}
\usepackage{graphicx,bbm,mathrsfs}
\usepackage{nicefrac}
\usepackage{bbm}
\usepackage{geometry}       
\usepackage{blkarray}       
\usepackage{multirow}       

\usepackage{jheppub}       

\usepackage{color}

\def\ie{{\it i.e.}}

\newcommand{\be}{\begin{equation}}  
\newcommand{\ee}{\end{equation}}  
\newcommand{\bea}{\begin{eqnarray}}  
\newcommand{\eea}{\end{eqnarray}}

\newcommand\lsim{\mathrel{\rlap{\lower4pt\hbox{\hskip1pt$\sim$}}
    \raise1pt\hbox{$<$}}}
\newcommand\gsim{\mathrel{\rlap{\lower4pt\hbox{\hskip1pt$\sim$}}
    \raise1pt\hbox{$>$}}}

\newcommand{\captionfonts}{\small}

\newcommand{\approptoinn}[2]{\mathrel{\vcenter{
  \offinterlineskip\halign{\hfil$##$\cr
    #1\propto\cr\noalign{\kern2pt}#1\sim\cr\noalign{\kern-2pt}}}}}

\usepackage{setspace}
\usepackage{array}

\newcolumntype{C}[1]{>{\centering\let\newline\\\arraybackslash\hspace{-20pt}}m{#1}}
\newcolumntype{L}[1]{>{\raggedleft\let\newline\\\arraybackslash\hspace{-20pt}}m{#1}}
\newcolumntype{R}[1]{>{\raggedright\let\newline\\\arraybackslash\hspace{-20pt}}m{#1}}

\makeatletter  
\long\def\@makecaption#1#2{%
  \vskip\abovecaptionskip
  \sbox\@tempboxa{{\captionfonts #1: #2}}%
  \ifdim \wd\@tempboxa >\hsize
    {\captionfonts #1: #2\par}
  \else
    \hbox to\hsize{\hfil\box\@tempboxa\hfil}%
  \fi
  \vskip\belowcaptionskip}
\makeatother   

\begin{document}

\hfill {\tt CERN-TH-2016-130}  

\vspace*{2.cm}

\begin{center}

\thispagestyle{empty}

{\Large\bf 
The correlation matrix of Higgs rates at the LHC
}\\[10mm]

\renewcommand{\thefootnote}{\fnsymbol{footnote}}

{\large Alexandre~Arbey$^{\,a,b,}$\footnote{alexandre.arbey@ens-lyon.fr}, 
\large Sylvain~Fichet$^{\,c,}$\footnote{sylvain@ift.unesp.br},
\large Farvah~Mahmoudi$^{\,a,b,}$\footnote{nazila@cern.ch}$^,$\textsuperscript{\ref{IUF}}, 
\large Gr\'egory~Moreau$^{\,d,}$\footnote{moreau@th.u-psud.fr}$^,$\addtocounter{footnote}{+1}\footnote{\label{IUF}Also Institut Universitaire de France, 103 boulevard Saint-Michel, 75005 Paris, France}}\\[10mm]

\addtocounter{footnote}{-6}

{\it
$^a$Univ Lyon, Univ Lyon 1, ENS de Lyon, CNRS, Centre de Recherche Astrophysique de Lyon UMR5574, F-69230 Saint-Genis-Laval, France \\
$^b$Theoretical Physics Department, CERN, CH-1211 Geneva 23, Switzerland \\
$^c$ICTP South American Institute for Fundamental Research, Instituto de Fisica Teorica 
Sao Paulo State University, Brazil \\
$^{d}$Laboratoire de Physique Th\'eorique, B\^at. 210, CNRS,
Universit\'e Paris-Sud 11 \\  F-91405 Orsay Cedex, France \\
}

\vspace*{12mm}

{\bf Abstract}
\end{center}
\noindent 

The imperfect knowledge of the Higgs boson decay rates and cross sections at the LHC constitutes a critical systematic uncertainty in the study of the Higgs boson properties.
We show that the full covariance matrix between the Higgs rates can be  determined  from  the most elementary sources of uncertainty by a direct application of probability theory.
We evaluate the error magnitudes and full correlation matrix on the set of Higgs cross sections and branching ratios at $\sqrt{s}=7$, $8$, $13$ and $14$~TeV, which are provided in ancillary files. The impact of
this correlation matrix on the global fits is illustrated with the latest $7$+$8$ TeV Higgs dataset.
\clearpage

\section{Introduction}

The measurement of the production and decay rates of the Higgs boson at the LHC \cite{ATLAS-disc,CMS-disc}  
has provided a new set of precision tests for the Standard Model (SM) of particle physics, giving access to the couplings of the Higgs to the other particles  of the SM. 
Such information on the Higgs couplings can in turn be recast into constraints on theories extending the SM.  The SM Higgs couplings being all fixed (now
that there is a measurement of the Higgs boson mass), any significant deviation would reveal the existence of a new physics lying beyond the Standard Model. 

This characterisation of the Higgs couplings relies on the comparison between the observed event numbers and those predicted by the SM  in a given detection channel. These theoretical predictions of the SM event rates are not infinitely precise. Rather, they are  plagued by a number of uncertainties, including the Parton Distribution Functions (PDF) uncertainties and the error inherent to the perturbative computation of amplitudes for the partonic processes.

The  uncertainties on the LHC Higgs production rates can reach about $10\%$ in relative magnitude. This is enough to substantially influence the outcome of the Higgs fits, so that the effect of these  uncertainties should be carefully  taken into account. As a matter of fact, the uncertainties on the cross sections have been found to have a non-negligible impact already on the fits of the $7+8$ TeV data (see e.g. Ref.~\cite{ATLAS-CMS-comb}). Besides, the statistical uncertainties will decrease with more data becoming available, hence the uncertainties on Higgs rates predictions are expected to become more and more important in the upcoming Higgs analyses.

The uncertainties on the SM Higgs production cross sections and partial decay widths are frequently taken into account in the global fits available in the literature. However, the \textit{correlations} among these uncertainties are poorly known, and are usually approximated or neglected. For example in the case of the 7+8 TeV CMS-ATLAS global fit of the Higgs couplings \cite{ATLAS-CMS-comb}, the correlation matrix among cross sections is approximated  with either $0\%$, $100\%$ or $-100\%$ elements. A similar approximation has been adopted in Ref.~\cite{Fichet:2015xla}. Yet, these Higgs rates correlations can in principle have a substantial impact on the fit \cite{Fichet:2015xla}, so that a complete treatment of the correlations is desirable.

The present paper is dedicated to address this shortcoming of the Higgs analyses, by providing a complete set of uncertainties on the Higgs rates at the LHC, including the correlations among all the rates.
Our study covers the uncertainties on \textit{inclusive} cross sections and branching ratios. No LHC-related experimental systematic uncertainties (on efficiencies or luminosity for example) are included.

We first show in Section~\ref{se:gen} that the complete correlation matrix of the Higgs rates can be directly obtained from a straightforward application of probability theory to the set of elementary sources of uncertainty. A necessary condition for this approach to work is that the relative magnitude of the elementary uncertainties be $\ll 100\%$ in order to allow error propagation, which will turn out to be a very good approximation. The subsequent formalism for error combinations is  described in Section~\ref{se:comb}. In Section~\ref{se:errors}, we then  enumerate all the elementary sources of  errors affecting the LHC Higgs rates predictions  and estimate their magnitudes, including the error on the Higgs mass determination. The full correlation matrix and combined errors on the  Higgs rates are given in Section~\ref{se:res}. These results are independent of the distributions associated to the elementary uncertainties. Finally in Section~\ref{se:appl}, we derive the analytical marginal likelihood including all correlations, using the fact that a central limit theorem is at work in presence of the large number of independent Higgs uncertainties
\cite{Fichet:2016qvx}. Fits of the full 7+8 Higgs dataset are also performed.



%

\section{Covariance and correlation matrix of expected Higgs rates }
\label{se:gen}

The likelihood associated to the Higgs events has in general the form 
\be
P({\rm\, Higgs\, data}\,|\,\theta,\delta) \equiv
{\cal L}(\theta,\delta)\,.
\ee
Here $\theta=(\theta_1,\theta_2,\ldots)$ denotes the  set of parameters of interest, which in the context of Higgs fits typically parametrises the deviations of the observed Higgs rates from the Standard Model predictions. 
The nuisance parameters $\delta=(\delta_1,\delta_2,\ldots)$  model any kind of systematic uncertainties. This paper is focused on the 
systematic uncertainties on the \textit{inclusive} Higgs rates, \ie~the SM predictions for production cross sections and decay widths with no phase space cuts.\footnote{Such systematic uncertainties are sometimes referred to as ``theoretical'' uncertainties, although this is only partially true as  some of them  are of experimental nature. 
Although this naming  provides a convenient contrast with the LHC-related experimental systematic uncertainties, such as the ones on efficiencies and luminosity, we will not use it to avoid any confusion. }
By definition, these uncertainties enter in the likelihood via the cross sections $\sigma_X$ and the decay widths $\Gamma_Y$,  so that the Higgs likelihood has in general the form
\be{\cal L}(\theta,\delta  ) \equiv {\cal L}[\theta,\sigma_X(\delta), \Gamma_Y(\delta) ]\,.
\ee
Here and below, we do not display the experimental nuisance parameters, which are irrelevant for this work.

Inferring information about the $\theta$ parameters necessitates to marginalise the unwanted parameters $\delta$, either through an integration in case of a Bayesian treatment, or a maximisation (\ie~profiling) in case of a  frequentist treatment, see  Ref.~\cite{Fichet:2015xla} for a review. In either case,  a ``prior'' distribution $\pi(\delta)$ is associated to the nuisance parameters. In practice, there can be  a large number of nuisance parameters $\delta$,  so that  marginalisation methods can be technically difficult to perform.  However, in the situation of small relative magnitudes (see 
Eq.\eqref{eq:def}), a simplification is available: 
 it is in principle possible  to combine all sources of uncertainty together \textit{before} marginalising.

Such combinations of uncertainties in both Bayesian and frequentist frameworks have been discussed in Ref.~\cite{Fichet:2015xla}, where it has been found that Bayesian combinations are both better defined and simpler than the frequentist ones. In the particular case of Gaussian priors,  Bayesian and frequentist combinations are found to be equivalent.
In the Bayesian case, the combination is defined by
\be \int \prod_n d \delta_n\, {\cal L}(\theta,\delta_n  ) \pi(\delta_n) \propto 
\int \prod_{X,Y} d \delta  \sigma_X  d \delta\Gamma_Y 
\,{\cal L}[\theta,\sigma_X^0+\delta \sigma_X , \Gamma_Y^0+\delta\Gamma_Y ] 
\bar\pi (\delta \sigma_X,\delta\Gamma_Y)
 \,,
 \label{eq:defL}
\ee
where the new dimensionful $\delta \sigma_X$, $\delta\Gamma_Y$ nuisance parameters encapsulate the combination of all the elementary sources of uncertainty, and $\sigma_X^0$ and $\Gamma_Y^0$ are the nominal values when no uncertainty is taken into account. 
 The practical interest of this combination is that the dimension of the integral on the right-handed side is lower than on the left-handed side, rendering the operation of marginalisation  more economic to achieve. 

The key point to correctly apply the method of preliminary combination lies in the determination of the combined prior $\bar\pi$. This might seem a very challenging task at first sight, as the $\bar\pi$ prior arises from a complicated combination of all elementary priors, and in general does not factorise.  However,  the \textit{covariance matrix} generated by $\bar \pi$, which is a central object for inference, is independent of the shape of the combined prior.  Moreover it is approximately independent of the elementary priors shape if elementary uncertainties have small relative magnitude (see Sec.~\ref{se:comb}).\footnote{The covariance matrix of the combined uncertainties is the second central moment of the $\bar \pi$ distribution. We recall that it is given by $V_{\alpha\beta}= \int d\delta_Z \delta_\alpha\, \delta_\beta \,  \bar \pi(\delta_Z) - \int d\delta_Z \delta_\alpha \,  \bar \pi(\delta_{Z})\,\, \int d\delta_{Z'} \delta_\beta \,  \bar \pi(\delta_{Z'})$. The combination of elementary uncertainties is prior-independent when they are linearly combined (see also \cite{Fichet:2015xla}).
 }
The covariance matrix arising from $\bar\pi$ can thus be determined without further reference to  prior shapes. This is what we are going to compute throughout this paper. 
Moreover, in case of a somewhat large number of sources of uncertainty with similar magnitudes, the shape of $\bar \pi$ automatically tends to a multivariate Gaussian by virtue of the Lyapunov central-limit theorem (see Ref.~\cite{Fichet:2016qvx}). In this case $\bar\pi$ is completely determined by the covariance matrix and the mean value - the latter  will always be set  to zero with no loss of generality  in our conventions. Finally, the  frequentist correlation matrix  matches the Bayesian one in case of  priors with Gaussian shapes. 

The goal of this work is to provide an accurate and consistent determination of the covariance matrix generated by $\bar\pi$, and the distribution of uncertainties on expected Higgs rates. In our analysis, for the combined uncertainty on decay rates,  we are going to work at the level of \textit{branching ratios} $\mathcal{B}_Y$ instead of partial widths.
 The error on partial widths are propagated into branching ratios following
\be
\mathcal{B}_Y=\mathcal{B}^0_Y+\delta\mathcal{B}_Y=\mathcal{B}^0_Y\bigg(1+ \frac{\delta\Gamma_Y}{\Gamma_Y} (1- \mathcal{B}^0_Y)-\sum_{Y'\neq Y} \mathcal{B}_{Y'}^0\frac{\delta\Gamma_{Y'}}{\Gamma_{Y'}}\bigg)\,. \label{eq:BR}
\ee
For further purpose,  it is convenient to put the  nuisance parameters under a standardised form, 
\be
\sigma_X=\sigma^0_X+\delta \sigma_X\equiv\sigma_X^0(1+\Delta_X\delta_X)\,,\quad \mathcal{B}_Y=\mathcal{B}^0_Y+\delta\mathcal{B}_Y\equiv {\cal B}_Y^0(1+\Delta_Y\delta_Y)\,, \label{eq:defdeltas}
\ee
where the normalised nuisance parameters $\delta_{X,Y}$ satisfy ${\rm E}[\delta_X]={\rm E}[\delta_Y]=0$, ${\rm Var}[\delta_X]={\rm Var}[\delta_Y]=1$, and the numbers $\Delta_X^2$, $\Delta_Y^2$ correspond to the relative variances for $\sigma_X$ and ${\cal B}_Y$. 
The interest of this notation is that the relative magnitudes of the uncertainties are explicitly factored out. In turn, the covariance matrix  ${\rm Cov}[\delta_X,\delta_{X'}]$ is precisely the \textit{correlation matrix} between the production modes. This applies similarly to the decay modes.\footnote{
An important conceptual remark is that the linear parametrisation in Eq.~\eqref{eq:defdeltas} requires the domain for $\delta$ to satisfy $\delta> -1/\Delta$ to forbid negative event numbers. For positive quantities, an exponential parametrisation $\sigma=\sigma_0 e^{\Delta\delta}$ is in principle much more natural. However, for the purpose of determining the covariance matrix,  using the linear parametrisation is enough.
}

Finally we join the production and decay labels by adopting a unified notation $Z=(X,Y)$, so that
\be
\delta_Z=(\delta_X,\delta_Y)\,,\quad \Delta_Z=(\Delta_X,\Delta_Y)\,.
\ee
Equation \eqref{eq:defL} is then equivalent to\be
\int d \delta_Z
\,{\cal L}[\theta,\sigma_X^0(1+\Delta_X\delta_X) , \Gamma_Y^0(1+\Delta_Y\delta_Y) ] 
\bar\pi (\delta_Z)\,.
\ee
The complete  correlation matrix of the expected Higgs rates generated by the $\bar \pi (\delta_Z) $ distribution is then given by
\be
\rho_{ZZ'}=\begin{pmatrix}
\rho_{XX'} & \rho_{YX'} \\
\rho_{XY'} & \rho_{YY'}
\end{pmatrix}\,,
\ee
and the complete  covariance matrix  is given by
\be
V_{ZZ'}=
\begin{pmatrix}
\Delta_X\Delta_{X'}\rho_{XX'} & \Delta_Y\Delta_{X'} \rho_{YX'} \\
\Delta_X\Delta_{Y'}\rho_{XY'} & \Delta_Y\Delta_{Y'} \rho_{YY'}
\end{pmatrix}\,.
\ee

We emphasise that the primary purpose of this paper is  to provide reliable covariance and correlation matrices associated with the expected Higgs rates. The choice of the exact shape for  $\bar\pi$ is left to the user -- although  a Gaussian shape is certainly well-motivated  \cite{Fichet:2015xla,Fichet:2016qvx}.

\section{Covariance matrix from combination of elementary uncertainties}
\label{se:comb}

Here we derive a general expression for the covariance matrix of quantities that are affected by arbitrary elementary uncertainties. 
This simple formalism will be applied to the Higgs rates in the next sections. The reader only interested in the final results for the Higgs can safely jump to Section~\ref{se:res}.

By definition the covariance relates two quantities at once. It is thus sufficient to  consider only the case of two quantities to obtain a fully general result. Let us consider two quantities $A$, $B$, subject to elementary sources of uncertainties respectively labelled by $n=(1,\ldots, p)$ and $n'=(1,\ldots, p')$. The nuisance parameters are denoted $\delta_n^A$, $\delta_{n'}^B$, so that $A\equiv A[\delta_1^A,\ldots,\delta_p^A]$, $B\equiv B[\delta_1^B,\ldots,\delta_{p'}^B]$. 

At this point, we make the crucial assumption that  the magnitude   of all relative uncertainties (\ie~the $\Delta$'s)  is small enough so that a Taylor expansion of $A$ and $B$ can be performed and stays valid up to $\delta=O(1)$. 
The most general form for the uncertainties is then
\be
A=A^0\bigg(1+\sum_{n=1}^p\delta_A^n\Delta_A^n+O\left((\Delta^n_A)^2\right)\bigg)\,,\quad B=B^0\bigg(1+\sum_{n'=1}^{p'}\delta^{n'}_B\Delta^{n'}_B+O\left((\Delta_B^{n'})^2\right)\bigg)\,.
\label{eq:def}
\ee
For the purpose of evaluating the covariance of $A$ and $B$, it is enough  to stop   the $\Delta$ expansion  at linear order. 
In the most general case, one should assume  correlations among all the sources of errors. 
The correlation matrix within the two groups of nuisance parameters $\delta^A_n$, $\delta^B_{n'}$ and the one between the two groups are respectively given by 
\be
\rho^{nm}_A={\rm Cov}[\delta^n_A,\delta^m_A]\,,\quad\rho^{n'm'}_B={\rm Cov}[\delta^{n'}_B,\delta^{m'}_B]\,,\quad\rho^{nn'}_{AB}={\rm Cov}[\delta^{n}_A,\delta^{n'}_B]\,.
\label{eq:corr}
\ee
In particular, if a given source of uncertainty $s$ affects both observables $A$ and $B$, one has $\rho_{AB}^{ss}=\pm1$.
The correlation matrices $\rho^{(nm)}_A$, $\rho^{(n'm')}_B$ and $\rho^{(nn')}_{AB}$ have respectively dimensions $p\times p$, $p'\times p'$ and $p\times p'$.

The next step is the combination of these elementary uncertainties. The global uncertainties on $A$ and $B$ are parametrised as
\be
A=A^0(1+\delta_A \Delta_A)\,,\quad B=B^0(1+\delta_B \Delta_B)\,.\label{eq:def_global}
\ee
Their correlation coefficient is denoted 
\be
\rho_{AB}={\rm Cov}[\delta_A,\delta_B]\,.\label{eq:corr_global}
\ee
Putting together Eqs.~\eqref{eq:def}--\eqref{eq:corr_global}, it follows that
\be
(\Delta_A)^2=\sum_{n,m} \rho^{nm}_A\Delta^{n}_A\Delta^{m}_A\,,\quad (\Delta_B)^2=\sum_{n',m'} \rho^{n'm'}_B\Delta^{n'}_B\Delta^{m'}_B\,,\label{eq:varAB}
\ee
and the correlation coefficient $\rho_{AB}$ is given by 
\be
\rho_{AB}\Delta_A\Delta_B= \sum_{n,n'} \rho^{nn'}_{AB}\Delta^{n}_A\Delta^{n'}_B\,. \label{eq:rhoAB}
\ee
Equations \eqref{eq:varAB} and \eqref{eq:rhoAB} fully specify the covariance matrix of $A$ and $B$, and thus determine entirely the errors on $A$ and $B$ induced by the elementary sources of error.
The covariance matrix of $(A, B)$ is expressed as 
\be
{\rm Cov}[A,B]=\begin{pmatrix}
(\Delta_A)^2 (A^0)^2 & \rho^{AB}\Delta_A\Delta_B\, A^0B^0 \\ \rho^{AB}\Delta_A\Delta_B\, A^0B^0 & (\Delta_B)^2 (B^0)^2
\end{pmatrix}
\,.
\ee

Of particular interest is the correlation matrix between the elementary uncertainties of the two observables, $\rho_{AB}^{nn'}$. From Eq.~\eqref{eq:corr_global}, it is clear that the more  coefficients $\rho_{AB}^{nn'}$ are close to one, the more  $A$ and $B$ get correlated. Similarly, the more coefficients $\rho_{AB}^{nn'}$ are close to zero, the more  $A$ and $B$ get uncorrelated. This shows explicitly  the competition between the uncertainties that correlate $A$ and $B$  and the ones that \textit{decorrelate} $A$ and $B$, which results in general in a non-trivial global correlation between $A$ and $B$. 
Another  observation is that the sources of uncertainty that are considered as elementary tend typically to be independent of each other.\footnote{ The property of independence follows rather naturally from the process of describing
systematic uncertainty as correctly as possible, as the more one delves into the origin of
uncertainty, the more its description becomes a set of elementary sources unrelated to
each other. The fact that most of elementary uncertainties are independent is not a crucial feature for  this paper, which is focussed on the combined covariance matrix. In contrast,  independence is very important to claim that the shape of the combined prior converges to a Gaussian (see \cite{Fichet:2016qvx}).
}
 Thus one can expect $\rho_A^{nm}=0$, $\rho_B^{nm}=0$ and $\rho_{AB}^{nn'}=0$ for $n\neq n'$. The case $n=n'$ corresponds instead to the same nuisance parameter entering in both $A$ and $B$, in which case the correlation is total, $\rho_{AB}^{nn'}=\pm 1$, as already mentioned.


In practice, the uncertainties that are not provided at the level of hadronic cross sections need to be propagated using a Taylor expansion with respect to $\Delta\ll 1$. This will be the case for most of the errors in this study. 
For a  cross section  $\sigma(Q)$ depending on a quantity $Q$ subject to an uncertainty described as
 $Q(\delta)=Q^0\times(1+\delta_Q\Delta_Q)$, the  error gets propagated through the cross section as 
\be\begin{split}
\sigma(Q)=\sigma(Q^0\times(1+\delta_Q\Delta_Q))&=\sigma(Q^0)\times \left(1+\frac{\partial \log\sigma}{\partial \log Q}\delta_Q\Delta_Q+O\left((\Delta_Q)^2\right)\right)\\&\equiv \sigma^0 \times(1+\delta_Q \Delta+O(\Delta^2))\,.
\end{split}
\ee
where $\sigma^0=\sigma(Q^0)$ and  the error on the cross section corresponds therefore to 
\be
\Delta=\frac{\partial \log\sigma}{\partial \log Q}\Delta_Q\,.
\label{eq:prop}
\ee
When necessary, the partial derivative in Eq.~\eqref{eq:prop} is obtained by varying the cross section with respect to the quantity $Q.$

\section{Inventory of the uncertainties on Higgs rates}
\label{se:errors}

At the LHC, the Higgs boson can be produced on-shell and  its decay products can be detected. The process of inclusive Higgs production followed by its decay is parametrised as
\be pp  \xrightarrow{X}( h \rightarrow Y)+\ldots \,\ee
where the ellipses denote extra states produced in association with the Higgs. 
The SM Higgs production mechanisms accessible at the LHC are {\it i)} gluon-gluon fusion (ggH), {\it ii)} vector boson
fusion (VBF), {\it iii)} associated production with an electroweak gauge
boson $V=W,Z$ (VH),  {\it iv)} associated production with a top quark pair $t\bar{t}$  (ttH), and {\it v)} associated production with a bottom quark pair $b\bar{b}$ (bbH). Moreover, the cross sections at different energies should be  considered separately  -- although a high correlation between the errors of a given cross section taken at different energies can be expected.

The production modes $X$ will be therefore taken in the following list,
\be
X=( X_{7~{\rm TeV}}, X_{8~{\rm TeV}}, X_{13~{\rm TeV}}, X_{14~{\rm TeV}} )
\ee
with
\be
X_{\sqrt{s}}=\{\textrm{ggH, VBF, ZH, WH,  ttH, bbH} \}_{\sqrt{s}}\,.
\ee
The main decay modes are 
\be
Y=\{ZZ, W^+W^-, \gamma\gamma, Z\gamma, gg, b\bar b , c\bar c, s\bar s , \tau\bar \tau, \mu\bar \mu \}\,.
\ee
The index of all Higgs rates $Z=( X,Y )=( X_{7~{\rm TeV}}, X_{8~{\rm TeV}}, X_{13~{\rm TeV}}, X_{14~{\rm TeV}} ,Y )$ takes  therefore its  values in a list of dimension 34.

Following Section \ref{se:gen}, the Higgs production cross sections are denoted as $\sigma_X$ and the partial decay width as $\Gamma_Y$.
The nuisance parameters  are written $\delta_X^n$ for a given cross section $\sigma_X$, and are respectively associated with relative magnitude $\Delta_X^n$. Therefore one has
\be
\sigma_X=\sigma_X^0\left(1+\sum_n\delta_X^n\Delta_X^n\right)\,.
\ee
The combined uncertainty is defined as 
\be
\sigma_X=\sigma_X^0\left(1+\delta_X\Delta_X\right)\,,
\ee
the correlation matrix is defined as 
\be
\rho_{XX'}={\rm Cov}[\delta_X,\delta_{X'}]\,,
\ee
and the complete covariance matrix as ${\rm Cov}[\sigma_X,\sigma_{X'}]=\sigma_X^0\sigma_{X'}^0\Delta_X\Delta_{X'}\rho_{XX'}\,.$
The notations are completely identical for branching ratios. For the latter, Eq.~\eqref{eq:BR} has to be used to translate partial width uncertainties into branching ratio uncertainties.

Having defined the formalism, we now turn to the inventory of all the elementary sources of uncertainty. Our aim is to specify the relative magnitudes of all the elementary systematic uncertainties $\Delta_Z$, as well as their possible correlations, $\rho_{Z}^{nm}$, $\rho_{Z'}^{n'm'}$, $\rho_{ZZ'}^{n n'}$, with $Z=(X,Y)$.
As discussed in the previous section, only the simultaneous variations of the rates are needed 
for the purpose of evaluating the correlation matrix among production modes. 
The details of these uncertainties have been discussed  in Ref.~\cite{Fichet:2015xla} and references therein, apart from the uncertainty from the Higgs mass which is, to the best of our knowledge, taken into account for the first time in this paper. We give here a brief summary of the Higgs rate uncertainties. 
\begin{itemize}

\item The experimental uncertainty on the  parton distribution functions are released by the PDF4LHC15 working group in Ref.~\cite{Butterworth:2015oua}. We use the PDF4LHC15\_100 error set, recommended for precision physics, which  includes a set of $100$ independent elementary nuisance parameters. These errors propagate into the cross sections with a relative magnitude $\Delta_X^{{\rm PDF},i}$, where $i=1\ldots100$.

\item The  uncertainties on $\alpha_s$, $m_t$, $m_b(\overline{MS})$, $m_c(3\, { \rm GeV})$ \cite{LHCHWG3}, and on the Higgs mass $m_h$ are given in Tab.~\ref{tab:param}. For the Higgs mass we combined in quadrature the statistical and systematic uncertainties reported in Ref.~\cite{Aad:2015zhl}.


\item The uncertainties from Effective Field Theory (EFT)  approximations in the ggH matrix elements are evaluated from Ref.~\cite{EFTorigin}. For the EFT approximation applied to the bottom quark, one has $\Delta^{{\rm EFT},b}_{\rm ggH}=4\%$ at the level of the ggH partonic cross section. Similarly, one has $\Delta^{{\rm EFT},V}_{\rm ggH}=4\%$ for the weak boson contributions.

\item The uncertainties from perturbative calculations of the QCD matrix elements are evaluated by varying the renormalisation scale $\mu_R$ within $[m_h/2,2m_h]$, with the central value $\mu_R^0=m_h$. One should bear in  mind that this choice for the variation is somewhat arbitrary.

\item The uncertainties in the DGLAP evolution of the PDFs (see Ref.\cite{PDG}, Section~9)  are evaluated by varying the factorisation scale $\mu_F$ within $[m_h/2,2m_h]$, with the central value $\mu_F^0=m_h$.

\end{itemize}

Finally, we need to specify the correlations among this set of elementary systematic uncertainties. Most of them are independent. The only non-trivial case is the one  of the $\mu_R$ and $\mu_F$ scale dependence. 
For both $\mu_R$ and $\mu_F$, some degree of correlation is expected among the production modes. To make sure we cover all the possibilities, we will consider two extreme cases, bearing in mind that the truth is somewhere in between. We consider
\footnote{
See Eq.~\eqref{eq:corr} for definitions. Let us recall that
$\rho_{X}^{\mu_R\mu_R}=\rho_{X}^{\mu_F\mu_F}=\rho_{X'}^{\mu_R\mu_R}=\rho_{X'}^{\mu_F\mu_F}=1$.}
 \\ \textit{\textit{i)} A fully independent case} where there is an independent $\mu_R^X$ and $\mu_F^X$ for each production mode, and any $\mu_R^X$ is independent of any $\mu_F^X$  so that
 \be \label{eq:Corri}
\rho_{X}^{\mu_R\mu_F}=\rho_{X'}^{\mu_R\mu_F}=\rho_{XX'}^{\mu_R\,\mu_R}=\rho_{XX'}^{\mu_F\,\mu_F}=\rho_{XX'}^{\mu_R\,\mu_F}=\rho_{XX'}^{\mu_F\,\mu_R}=0\quad {\rm for \,\, any }\,\, (X,X')\,. 
\ee
\\ \textit{\textit{ii)} A fully correlated case} where a universal $\mu_R$ and $\mu_F$ is assumed for all production modes, and $\mu_R$ and $\mu_F$ are further assumed to be $100\%$ correlated so that there is a single nuisance parameter $\mu_R=\mu_F\equiv \mu$. The correlation matrices are therefore\footnote{Notice that having $\rho_{X}^{\mu_R\mu_F}=\rho_{X'}^{\mu_R\mu_F}=\rho_{XX'}^{\mu_R\,\mu_R}=1$ is a sufficient condition to fix the three other correlations to one. }
 \be \label{eq:Corrii}
\rho_{X}^{\mu_R\mu_F}=\rho_{X'}^{\mu_R\mu_F}=\rho_{XX'}^{\mu_R\,\mu_R}=\rho_{XX'}^{\mu_F\,\mu_F}=\rho_{XX'}^{\mu_R\,\mu_F}=\rho_{XX'}^{\mu_F\,\mu_R}=1\quad {\rm for \,\, any }\,\, (X,X')\,.
\ee

\begin{table}
\center
\begin{tabular}{|c|c|c|c|c|c|}
\hline
Input & $\alpha_s$ & $m_t$ & $m_b (\overline{\rm MS})$ & $m_c(3\, { \rm GeV})$ & $m_h$ \\
\hline
Mean value & $0.118$ & $172.5$ GeV & $4.18$ GeV & $0.986 $ GeV & $ 125.09 $ GeV \\

$\Delta$ & $1.27\%$  & $0.58\%$ & $0.72\%$ & $2.6\%$ & $0.19\%$ \\

Ref. & \cite{Butterworth:2015oua} & \cite{LHCHWG3} & \cite{LHCHWG3} & \cite{LHCHWG3}  & \cite{Aad:2015zhl} \\
\hline
\end{tabular}

\caption{Parametric uncertainties affecting the Higgs production and decay rates.
\label{tab:param}}
\end{table}
{\renewcommand{\arraystretch}{2}

\section{Total error and correlation matrix of the Higgs rates at the LHC} 
\label{se:res}

Having determined the magnitude of the elementary uncertainties, we can readily apply the general analysis of Section~\ref{se:comb}. 
We evaluate numerically the simultaneous variations of all Higgs rates with respect to the elementary uncertainties listed in Section \ref{se:errors}. 
From these  partial derivatives (see Eq.\eqref{eq:prop}), one obtains the $\Delta$'s and $\rho$'s that enter in the combination formulas Eqs.~\eqref{eq:varAB}, \eqref{eq:rhoAB}. This fully determines the covariance matrix of the Higgs rates.
 
 To compute the cross sections we use \texttt{SusHi v1.5.0}~\cite{sushi} for ggH and bbH,  modified versions of \texttt{VV2H v1.10}~\cite{Han:1992hr,SpiraCodes} and \texttt{V2HV v1.10}~\cite{Han:1991ia,SpiraCodes} interfaced with LHAPDF~\cite{Buckley:2014ana} for VBF and VH, and an NLO version of \texttt{HQQ}~\cite{SpiraPrivate,SpiraCodes} for ttH, based on Ref.~\cite{Dittmaier:2014sva}. For the ZH mode we also included the $gg\rightarrow ZH$ contribution, computed with \texttt{VH@NNLO v1.2.1}}~\cite{Brein:2012ne}.  The decay widths have been computed using \texttt{HDECAY v6.51} \cite{Djouadi:1997yw}. 
The nominal values for $\sigma_X^0$ and $\Gamma_Y^0$ are given in Tabs.~\ref{tab:XS} and \ref{tab:decays}, respectively. These generators have been chosen in order to provide reproducible results. All hypotheses and cuts used to generate these cross sections are in principle publicly available.


The complete $34\times 34$ correlation matrix $\rho_{ZZ'}$  and 
the vector  of relative magnitudes $\Delta_Z=(\Delta_{X_{7\,{\rm TeV}}}, \Delta_{X_{8\,{\rm TeV}}}, \Delta_{X_{13\,{\rm TeV}}}, \Delta_{X_{14\,{\rm TeV}}},\Delta_Y)$, all in both scale correlation cases $i)$ and $ii)$,  are included in ancillary files 
attached to the arXiv version of this manuscript. 
The relative magnitudes $\Delta_Z$ are shown  in Tab.~\ref{tab:vectors}.
The  $7$, $8$, $13$, $14$ TeV blocks of the correlation matrices are shown in the Appendix, in Tabs.~\ref{tab:rho7} to \ref{tab:rho14ii}. The correlations of cross sections between various energies, which are encoded in the non-diagonal blocks of the complete $\rho_{ZZ'}$ matrix given in the ancillary files, can deviate by $\sim 10\%$ with respect to the correlations of cross sections at same energy.\footnote{
The choice of generators is expected to affect only marginaly the correlation matrices. We checked that results given by, for example,  \texttt{HIGLU v4.34}~\cite{Spira:1995rr,Spira:1996if} and \texttt{SusHi} give very similar uncertainty on the ggF cross-section.  The discrepancy between the two estimations of the error is of order 0.1\%.  Moreover, the central values from \texttt{HIGLU} and \texttt{SusHi}  are well compatible within these uncertainties. These cross-sections are in good agreement with NNLO values of the LHCXSWG. 
}

\begin{table}
\center
\begin{tabular}{|c|c|c|c|c|c|c|}
\hline
Process $X$ & ggH & VBF & ZH & WH  & tth  & bbh \\
\hline
Calculation & \texttt{SusHi} \cite{sushi} & \texttt{VV2H} \cite{SpiraCodes} & \texttt{V2HV} \cite{SpiraCodes} & \texttt{V2HV}  & \texttt{HQQ} \cite{SpiraPrivate}& \texttt{SusHi}   \\
Order & NNLO & NLO & NLO & NLO & NLO & NNLO \\
\hline
$\sigma_X^0(7\, {\rm TeV})$ [fb]  & 14.5  & 1.27  & 0.34 & 0.60  & 0.089   & 0.171  \\

$\sigma_X^0(8\, {\rm TeV})$ [fb]  & 18.5 & 1.63 & 0.424 & 0.729 & 0.134   & 0.223 \\

$\sigma_X^0(13\, {\rm TeV})$ [fb]  & 42.2 & 3.87 & 0.890 & 1.42 & 0.516  &  0.545  \\

$\sigma_X^0(14\, {\rm TeV})$ [fb]  & 47.6 & 4.39 & 0.991 & 1.56 & 0.624 & 0.618  \\
\hline
\end{tabular}

\caption{Cross sections. \label{tab:XS}}
\end{table}

\begin{table}
\center
\begin{tabular}{|c|c|c|c|c|c|c|c|c|c|c||c|}
\hline
Channel $Y$ & $WW$  & $ZZ$ & $\gamma\gamma$ & $Z\gamma$  & $gg$  & $b\bar b$ & $c\bar c$ & $s\bar s$ & $\tau\bar \tau$ & $\mu\bar\mu$ & Total \\
 $\Gamma_Y^0$ [MeV] &   0.89 & 0.11 & 0.009 & 0.006 & 0.34 & 2.38 & 0.12 &  0.001  & 0.28 & 0.0009   & 4.10 \\ \hline
\end{tabular}

\caption{Higgs decay widths. \label{tab:decays}}
\end{table}
{\renewcommand{\arraystretch}{2}

\begin{table}
\be \nonumber
\renewcommand*{\arraystretch}{1.5}
\begin{array}{C{1.cm}C{0.8cm}*{6}{C{2.cm}}L{0.2cm}}
 & & ggH & VBF & ZH & WH  & ttH  & bbH  
  \\
 $\Delta_{X_{7\,{\rm TeV}}}=$ & $\Big($ &  12.3 (12.1) & 2.0 & 2.5 (2.2) & 2.6\,(2.4) & 13.7 (18.9) & 11.8 (10.6) &  $\Big)$ \\
 \end{array}
\ee
\be \nonumber
\renewcommand*{\arraystretch}{1.5}
\begin{array}{C{1.cm}C{0.8cm}*{6}{C{2.cm}}L{0.2cm}}
 $\Delta_{X_{8\,{\rm TeV}}}=$ & $\Big($ &  12.3 (11.9) & 2.0 & 2.4 (2.1) & 2.6 (2.2) & 13.6 (18.9) & 11.6 (10.2) &  $\Big)$ \\
 \end{array}
\ee
\be \nonumber
\renewcommand*{\arraystretch}{1.5}
\begin{array}{C{1.2cm}C{0.6cm}*{6}{C{2.cm}}L{0.2cm}}
 $\Delta_{X_{13\,{\rm TeV}}}=$ & $\Big($ &   12.3 (11.2) & 2.1 (2.2) & 2.4 (1.8) & 2.5 (2.0) & 13.6 (19.0) & 11.0 (8.9) &  $\Big)$ \\
 \end{array}
\ee
\be \nonumber
\renewcommand*{\arraystretch}{1.5}
\begin{array}{C{1.2cm}C{0.6cm}*{6}{C{2.cm}}L{0.2cm}}
 $\Delta_{X_{14\,{\rm TeV}}}=$ & $\Big($ &  12.3 (11.1) & 2.1 (2.2) & 2.5 (1.8) & 2.6 (1.9) & 13.6 (19.0) & 10.9 (8.7) &  $\Big)$ \\
 \end{array}
\ee


\be \nonumber
\renewcommand*{\arraystretch}{1.5}
\begin{array}{*{13}{C{0.9cm}}}
 & & $WW$ & $ZZ$ & $\gamma\gamma$ & $Z\gamma$  & $gg$  & $b\bar b$ & $c\bar c$ & $s\bar s$ & $\tau\bar \tau$ & $\mu\bar\mu$ 
  \\
 $\Delta_Y=$ & $\Big($ & 2.1 & 2.4 & 1.2 & 1.8 & 3.9 & 1.3 & 5.4 & 2.4 & 1.3 & 1.3 & $\Big)$ \\
 \end{array}
\ee

\caption{Magnitude of the relative uncertainties  of the expected Higgs  production and decay rates, assuming  scale correlations $i)$ or $ii)$ (shown in parenthesis), as defined in Section~\ref{se:errors}.  The magnitudes are given in percent. \label{tab:vectors}}
\end{table}
%
%
%


\subsubsection*{On the correlations among ggH and ttH production modes}

Our result on the correlation between the ggH and ttH modes depends strongly on the assumptions for scale correlations. 
The correlations we obtain are either very small or positive.  
In case $i)$ we obtain 
\begin{eqnarray}
 \rho_{\rm ggH_{7\rm TeV}\, ttH_{7\rm TeV}}=-1.2\%\,,\quad\rho_{\rm ggH_{8\rm TeV}\, ttH_{8\rm TeV}}=-1.3\%\,,  \nonumber \\
 \quad \rho_{\rm ggH_{13\rm TeV}\, ttH_{13\rm TeV}}=-1.6\%\,,\quad \rho_{\rm ggH_{14\rm TeV}\, ttH_{14\rm TeV}}=-1.5\%\,,   
 \label{rho_ggh_tth_i}
\end{eqnarray} 
while for case $ii)$ we get 
\begin{eqnarray}
 \rho_{\rm ggH_{7\rm TeV}\, ttH_{7\rm TeV}}=82\%\,,\quad\rho_{\rm ggH_{8\rm TeV}\, ttH_{8\rm TeV}}=82\%\,,  \nonumber \\
 \quad \rho_{\rm ggH_{13\rm TeV}\, ttH_{13\rm TeV}}=80\%\,,\quad \rho_{\rm ggH_{14\rm TeV}\, ttH_{14\rm TeV}}=80\%\,.   
  \label{rho_ggh_tth_ii}
\end{eqnarray} 
These numbers arise from a non-trivial competition among  the various sources of uncertainties. Indeed, the
 PDF uncertainties and the $\alpha_s$ error tend to induce a negative correlation, while the $m_h$ and $m_t$  errors induce a positive correlation.   
 Without the contribution from scale errors, the total correlations would be $O(-10\%)$, which is in reasonable agreement with the result of Ref.~\cite{Butterworth:2015oua} ({\it c.f.} 
 Table~3) [see also Ref.~\cite{LHCHWGII} ({\it c.f.} 
 Table~10)].\footnote{The correlation of $\sim -60\%$ between ggH and ttH reported in Ref.~\cite{CMS-NOTE-2011-005} has been revised.} 
 However it turns out that the scale errors are large enough to completely change this value. 
 In case $i)$, where scale errors are independent,  the ggH-ttH correlation changes to $O(-1\%)$ (see Eq.~\eqref{rho_ggh_tth_i}). 
   In case $ii)$,  where scale errors are $100\%$ correlated, the ggH-ttH  correlation becomes $O(80\%)$ (see Eq.~\eqref{rho_ggh_tth_ii}).\footnote{In addition to these extreme cases, one can also consider an intermediate one, close to $i)$, where the scales $\mu_R^X$ and $\mu_F^X$ are $100\%$ correlated for a given cross-section $X$, but are independent between each different  cross-sections. This corresponds to the configuration
 $ \label{eq:Corri}
\rho_{X}^{\mu_R\mu_F}=\rho_{X'}^{\mu_R\mu_F}=1,~\rho_{XX'}^{\mu_R\,\mu_R}=\rho_{XX'}^{\mu_F\,\mu_F}=\rho_{XX'}^{\mu_R\,\mu_F}=\rho_{XX'}^{\mu_F\,\mu_R}=0$ for any $(X,X')$. 
The ggF-ttH correlations are then 
\begin{eqnarray}
 \rho_{\rm ggH_{7\rm TeV}\, ttH_{7\rm TeV}}=-0.88\%\,,\quad\rho_{\rm ggH_{8\rm TeV}\, ttH_{8\rm TeV}}=-0.88\%\,,  \nonumber \\
 \quad \rho_{\rm ggH_{13\rm TeV}\, ttH_{13\rm TeV}}=-1.3\%\,,\quad \rho_{\rm ggH_{14\rm TeV}\, ttH_{14\rm TeV}}=-1.3\%\,.
 \label{rho_ggh_tth_int}
\end{eqnarray} 
These numbers turn out to be close to those of Eq.~\eqref{rho_ggh_tth_i}.
}
   The $\rho_{\rm ggH\, ttH}$ case is a good example of a non-trivial combination of uncertainties, 
   that requires  the formalism presented in Sec.~\ref{se:comb} in order to be treated correctly.

\section{Application to 7+8 TeV Higgs data}
\label{se:appl}

\begin{figure}[t]
\begin{picture}(400,200)
\put(00,0){
\includegraphics[width=7.5cm]{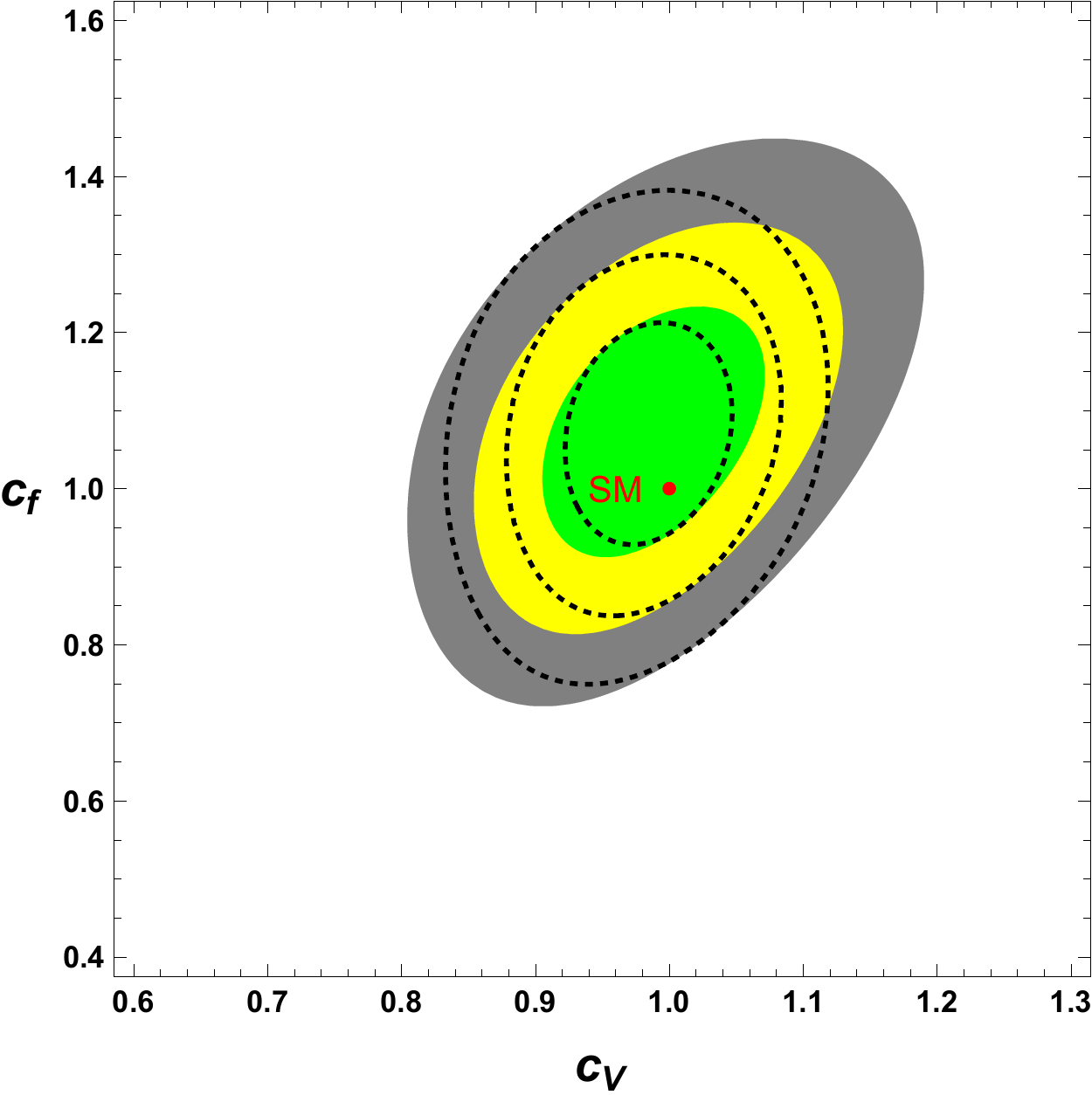}\quad
\includegraphics[width=7.5cm]{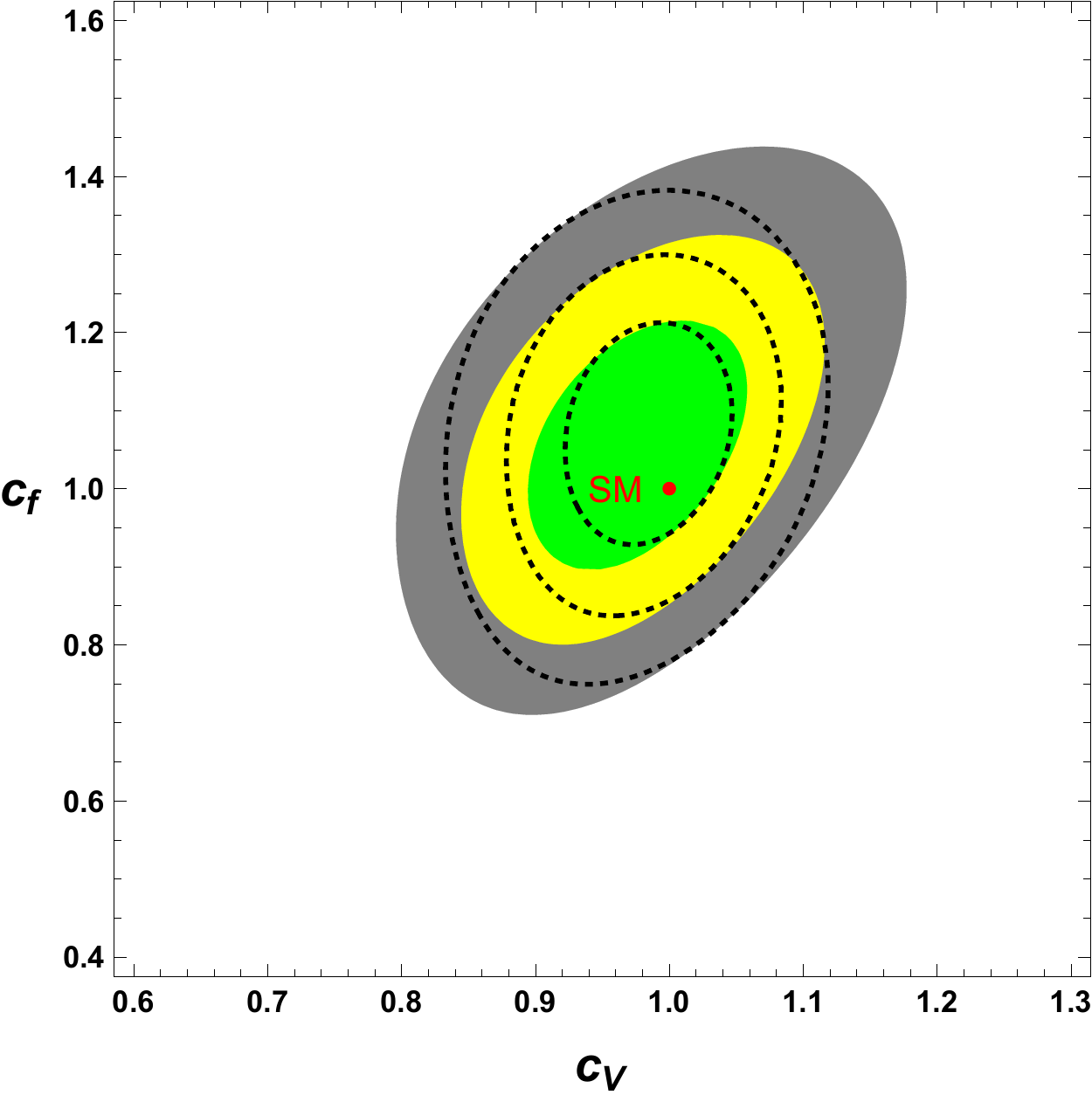}}
\put(40,50){{\color{blue} Bayesian marginalisation}}
\put(40,35){{\color{blue} -- {\it Scale correlations $i)$ } --}}
\put(270,50){{\color{blue} Bayesian marginalisation}}
\put(270,35){{\color{blue} -- {\it Scale correlations $ii)$ } --}}
\end{picture}
\caption{
Best-fit regions in the $c_V-c_f$ plane obtained from Bayesian marginalisation.
The $68\%$, $95\%$ and $99\%$ credible regions when  theoretical uncertainties are taken into account
are represented respectively  by the green, yellow and grey domains. 
The dashed contours illustrate the case without theoretical uncertainties. The SM prediction is shown by a red point.
The two extreme cases for renormalisation and factorisation scale correlations, leading to the two plots,  are described in Eqs.~\eqref{eq:Corri}, \eqref{eq:Corrii}.
}\label{fig:CombFull}
\end{figure}
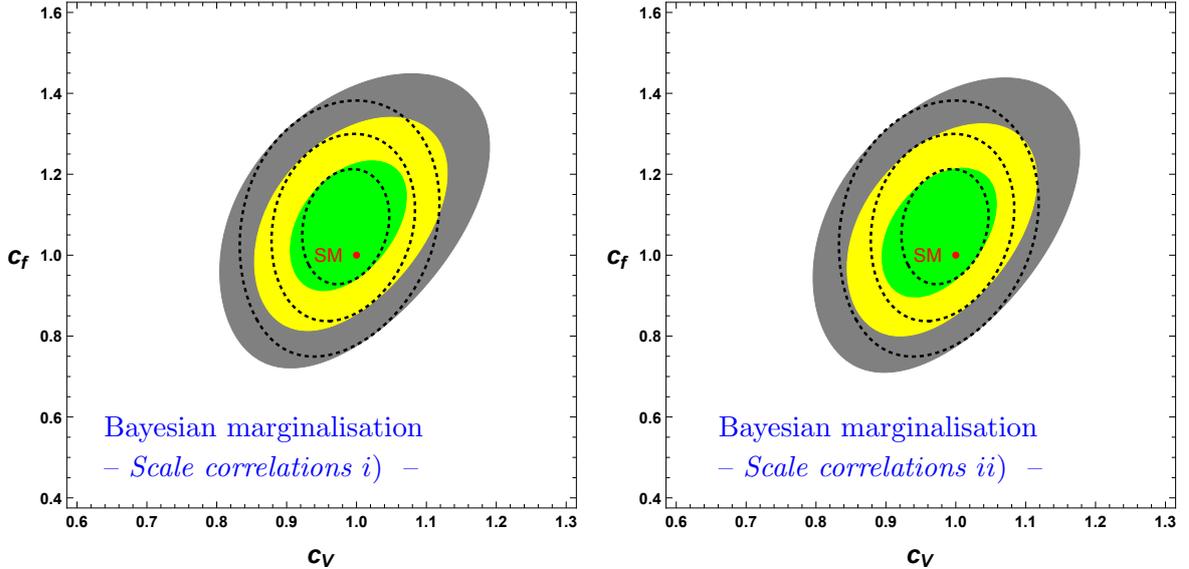

In this section we compute and display the simplified marginal likelihood, following closely the derivation of Ref.~\cite{Fichet:2016qvx}.\footnote{The derivation of the simplified marginal likelihood consists of steps of error propagation and combination, and uses the fact that a central limit theorem applies to the combined errors.}  Here we perform a Bayesian marginalisation, \ie~an integration over the nuisance parameters $\delta_{Z}$, with a multivariate normal prior  whose covariance matrix is the $\rho_{ZZ'}$ matrix obtained through the previous sections.  
This marginal likelihood is analytical and involves explicitly $\rho_{ZZ'}$ and the relative magnitudes $\Delta_Z$. 
One introduces the observed and expected signal strengths $\hat \mu_I$ and  $\mu_I(\theta)$, the latter being a function of a set of parameters of interest $\theta$. The absolute statistical error on the signal strength $\mu_I$ is written  $\Delta\mu_I$.
The experimental efficiencies for a cross section $X$ in a detection channel $I$ are denoted $\epsilon_X^I$.\footnote{The signal strengths are defined as $\hat \mu_I=\hat N_I / N_I^{\rm SM}$, $ \mu_I (\theta)= N_I^{\rm BSM}(\theta) / N_I^{\rm SM}$, where $\hat N_I$ is the observed event number in channel $I$, $N_I^{\rm SM}=\sum_X\epsilon_X^{\textrm{SM}\,I}\sigma_X {\cal B}_{Y(I)}$ is the expected event number in the Standard Model, and similarly $N^{\rm BSM}(\theta)_I$ is the expected event number in a modelisation of physics beyond the SM with parameters $\theta$. In this section,
 it is assumed selection efficiencies such that, $\epsilon_X^{\textrm{BSM}\,I}=\epsilon_X^{\textrm{SM}\,I}\equiv \epsilon_X^I$, following Refs. \cite{Moreau:2012da,Dumont:2013wma}. 
This relation is a good approximation when using Higgs signal strengths \cite{Dumont:2013wma},  is exact in the absence of higher-order derivative operators, and allows us  to conveniently estimate the $N^{\rm BSM}(\theta)_I$ event numbers.}

The marginal likelihood is given by
\be
\bar L (\theta)= L^{\rm stat}(\theta) L^{\rm sys}(\theta) =L^{\rm stat}(\theta)\, \frac{1}{\sqrt{|\eta \rho+1|}} \exp\left(  \frac{1}{2}
 \xi \cdot \Big( \eta  +\rho^{-1}   \Big)^{-1} \cdot  \xi
  \right)\,, \label{eq:Ltildegen}
\ee
with
\be
\xi_Z=\sum_I \mu_I(\theta) \frac{\hat \mu_I-\mu_I(\theta)}{(\Delta\mu_I)^2} \Delta^I_Z\,,
\ee
\be
\eta_{ZZ'}=\sum_I  \frac{\left(\mu_I\left(\theta\right)\right)^2}{(\Delta\mu_I)^2} (\Delta^I_Z)^2 \delta_{ZZ'}\,,
\ee
and
\be
\Delta^I_Z=\begin{pmatrix}
\epsilon_X^I\sigma_X\Delta_X \left(\sum_{X'} \epsilon_{X'}^I\sigma_{X'}\right)^{-1} \\
\Delta_Y^I 
\end{pmatrix}\,,
\ee
where  $\delta_{ZZ'}$ is the Kronecker delta, and $\Delta_Y^I$ is the branching ratio uncertainty of the final state $Y$ selected in the channel $I$.  The $L^{\rm stat}(\theta)$ term is the Poisson likelihood containing the statistical error only.
The $L^{\rm sys}(\theta)$ term encodes the  effects of the systematic uncertainties.  In particular, it contains all the correlations among signal strengths induced by the systematic uncertainties.

This computation is in fact slightly simpler than the general case presented in Ref.~\cite{Fichet:2016qvx}. Although the number of elementary uncertainties is large, they get combined in only a few nuisance parameters $\delta_{X,Y}$ at the level of cross sections and branching ratios, which have been checked in Ref.~\cite{Fichet:2015xla} to approximately follow a multivariate normal distribution, as expected from the central limit theorem. Because the number of nuisance parameters is smaller than the number of available Higgs channels, it is convenient to marginalise over these nuisance parameters instead of propagating the errors  up to the event numbers. 

As a final application, we present a global fit of the full set of $7+8$ TeV data, following closely Ref.~\cite{Fichet:2015xla}.
For  ATLAS data, 
the diphoton final state results are taken from Ref.~\cite{A-diphoton}, 
the $ZZ$ channel is from Ref.~\cite{A-ZZ}, 
the $WW$ channel from Ref.~\cite{A-WW}, 
the $b \bar b$ from Ref.~\cite{A-bb} and 
the $\tau\bar \tau$ from Ref.~\cite{A-tautau}.
The combined channels are studied in Ref.~\cite{ATLASfit}.
For  CMS data,
the diphoton final state has been presented in Ref.~\cite{C-diphoton}, 
the $ZZ$ channel measurements are provided in Ref.~\cite{C-ZZ}, 
the $WW$ ones in Ref.~\cite{C-WW},
the $b \bar b$ in Ref.~\cite{C-bb} and 
the $\tau\bar \tau$ in Ref.~\cite{C-tautau} (see also the combined channel analysis~\cite{CMSfit}).

 As a simple model of potential new physics effects, the deviations of the expected signal strengths from the SM are parametrised  as
\begin{eqnarray} 
{\cal L}_H  & = &  
 \ c_V \left( \ g_{hWW} \ h \ W_{\mu}^+ W^{-\mu} +  \ g_{hZZ} \ h \  Z_{\mu}^0 Z^{0 \mu}\right) \nonumber \\
  & & - c_f \left( \ y_t \ h \ \bar t_L t_R \  +  y_b \ h \ \bar b_L b_R \ +  y_c \ h \ \bar c_L c_R \ +  y_\tau \ h \ \bar \tau_L \tau_R \right) + \ {\rm h.c.}  
\label{Eq:LagEff}
\end{eqnarray} 
where $y_{t,b,c,\tau}$ are the SM Yukawa coupling constants (in mass eigenbasis), 
the subscript $L/R$ indicates the fermion chirality,  
$v$ is the Higgs vacuum expectation value, 
$g_{hWW} = 2M^2_W/v$ and $g_{hZZ} = M^2_Z/v$ are the electroweak gauge boson couplings.
 The $c_{V,f}$ parameters are defined  such that the limiting case $c_{V,f}\to 1$ corresponds to the SM. 

The Bayesian regions at $68,95,99\%$ credibility level are presented in the $c_V-c_f$ plane in Fig.~\ref{fig:CombFull}, assuming flat prior for $c_{V,f}$. The shapes of the regions are rather similar between the two extreme cases for renormalisation/factorisation scale correlations. The main difference between the two extreme cases  turns out to be the shift of the best-fit regions with respect to the fit without Higgs rate uncertainties (shown with dotted regions in Fig.~\ref{fig:CombFull}). This discrepancy between the two extreme cases of scale correlations illustrates the 
importance of properly including the  correlation matrices in the analyses of the Higgs rates. 
The impact of correlations can also be seen by comparing Fig.~\ref{fig:CombFull} with
the best-fit regions of Fig.~6 from Ref.~\cite{Fichet:2015xla},
 where  schematic correlations have been used.

As the LHC will accumulate more and more data on the Higgs processes, 
 the statistical error bars will decrease, letting the uncertainties on the Higgs rates be among the dominant ones.  Therefore, the correct treatment of these uncertainties  
  is expected to become more and more crucial for the global Higgs fits.

\section{Conclusion} 
\label{se:conc}

We have evaluated  the full covariance matrix 
of the expected Higgs production and decay rates, through a combination of their elementary sources of uncertainty.  Our calculation follows from a direct application of probability theory, and is completely consistent with  the framework of Bayesian statistics. The obtained covariance matrix is  prior-independent, and is also consistent with a frequentist combination of Gaussian elementary uncertainties. A frequentist combination of non-Gaussian uncertainties is prior-dependent and beyond the scope of this work.

We provide  the error magnitudes and correlation matrices on the set of Higgs cross sections and branching ratios at $\sqrt{s}=7$, $8$, $13$ and $14$~TeV. Most of the elementary uncertainties are evaluated via error propagation in the  
\texttt{SusHi}, \texttt{VV2H}, \texttt{V2HV}, \texttt{HQQ} and \texttt{HDECAY} codes. This error propagation is equivalent to truncating the expansion in the relative magnitude of the errors ($\Delta$) at first order, which we systematically checked to be a good approximation. Our set of elementary uncertainties includes the error on the Higgs mass determination by CMS and ATLAS. 

The error magnitudes and the correlation matrices are provided under various formats   with the  arXiv version of this paper,  
and can be readily used for further global fits of the Higgs data. 
The lack of knowledge about the correlations among the elementary uncertainties coming from the various renormalisation and factorisation scales has been taken into account by considering two extreme cases for these correlations.
In these extreme cases, we find for example either a very small or a large positive correlation between the ggH and ttH rates. 

Finally we present global fits of the latest $7+8$ TeV dataset, showing the impact of the full Higgs rate covariance matrix   on the Higgs coupling determination. In doing so we use a simplified likelihood framework \cite{Fichet:2016qvx}, relying  on the fact that the distribution of the combined nuisance parameters is approximately Gaussian.

\section*{Acknowledgements} 

We would like to thank Michael Spira and Gavin Salam for useful discussions. The work of S.F. was supported by the
S\~ao Paulo Research Fundation (FAPESP) under grant 2014/21477-2. G.M. is supported by the ERC advanced grant Higgs@LHC and the European
Union FP7 ITN Invisibles.
\\
\\
\\
\\
\\
\appendix

\clearpage

\section{Blocks of the Higgs rate correlation matrix}
\label{app:BRs}

Here we display the $7$, $8$, $13$, and $14$ TeV blocks of the correlation matrix $\rho_{ZZ'}$, assuming either scale correlations $i)$   or $ii)$, as defined in Eqs.~\eqref{eq:Corri} and \eqref{eq:Corrii}.

\begin{table}[h]
\be
\small
\begin{blockarray}{*{17}{c<{\hspace{0pt}}}}
 \text{ggH} & \text{VBF} & \text{ZH} & \text{WH} & \text{ttH} & \text{bbH} & \,\,WW & ZZ & \gamma \gamma  &  Z \gamma &  gg & b\bar{b} & c\bar{c} & s\bar{s} & \tau \bar{\tau } & \mu \bar{\mu } &  \\
 \begin{block}{(*{15}{c<{\hspace{-10pt} }}c)c}
 100 & 8 & 0 & -1 & -1 & 3 & 3 & 2 & 11 & 5 & 18 & -11 & -4 & -16 & 11 & 11 & \text{ggF} \\
 8 & 100 & 49 & 36 & -9 & -25 & -3 & -4 & 13 & 0 & 23 & -9 & -4 & -18 & 16 & 16 & \text{VBF} \\
 0 & 49 & 100 & 68 & 0 & -13 & -18 & -19 & 6 & -15 & 12 & 6 & 0 & -3 & 17 & 17 & \text{ZH} \\
 -1 & 36 & 68 & 100 & 2 & -4 & -17 & -18 & 5 & -14 & 11 & 6 & 0 & -3 & 16 & 16 & \text{WH} \\
 -1 & -9 & 0 & 2 & 100 & 10 & -4 & -4 & -1 & -3 & 0 & 3 & 1 & 2 & 2 & 2 & \text{ttH} \\
 3 & -25 & -13 & -4 & 10 & 100 & -7 & -7 & -4 & -7 & 10 & 2 & -5 & -17 & -1 & -1 & \text{bbH} \\
 3 & -3 & -18 & -17 & -4 & -7 & 100 & 100 & 56 & 99 & 36 & -86 & -15 & -29 & 10 & 10 & WW \\
 2 & -4 & -19 & -18 & -4 & -7 & 100 & 100 & 49 & 98 & 31 & -83 & -15 & -29 & 2 & 2 & ZZ \\
 11 & 13 & 6 & 5 & -1 & -4 & 56 & 49 & 100 & 67 & 77 & -82 & -11 & -15 & 88 & 88 & \gamma \gamma  \\
 5 & 0 & -15 & -14 & -3 & -7 & 99 & 98 & 67 & 100 & 46 & -91 & -16 & -28 & 24 & 24 & \text{ $Z\gamma$} \\
 18 & 23 & 12 & 11 & 0 & 10 & 36 & 31 & 77 & 46 & 100 & -73 & -20 & -70 & 70 & 70 & gg \\
 -11 & -9 & 6 & 6 & 3 & 2 & -86 & -83 & -82 & -91 & -73 & 100 & -1 & 46 & -48 & -48 & b\bar{b} \\
 -4 & -4 & 0 & 0 & 1 & -5 & -15 & -15 & -11 & -16 & -20 & -1 & 100 & 25 & -4 & -4 & c\bar{c} \\
 -16 & -18 & -3 & -3 & 2 & -17 & -29 & -29 & -15 & -28 & -70 & 46 & 25 & 100 & 1 & 1 & s\bar{s} \\
 11 & 16 & 17 & 16 & 2 & -1 & 10 & 2 & 88 & 24 & 70 & -48 & -4 & 1 & 100 & 100 & \tau \bar{\tau } \\
 11 & 16 & 17 & 16 & 2 & -1 & 10 & 2 & 88 & 24 & 70 & -48 & -4 & 1 & 100 & 100 & \mu \bar{\mu } \\
 \end{block}
\end{blockarray}
\nonumber
\ee
\caption{Correlation matrix between the expected Higgs  production and decay rates at  $\sqrt{s}=7$ TeV, expressed in percent, assuming the scale correlations $i)$. \label{tab:rho7} }
\end{table}

\begin{table}
\be
\small
\begin{blockarray}{*{17}{c<{\hspace{0pt}}}}
 \text{ggH} & \text{VBF} & \text{ZH} & \text{WH} & \text{ttH} & \text{bbH} & \,\,WW & ZZ & \gamma \gamma  &  Z \gamma &  gg & b\bar{b} & c\bar{c} & s\bar{s} & \tau \bar{\tau } & \mu \bar{\mu } &  \\
 \begin{block}{(*{15}{c<{\hspace{-10pt} }}c)c}
 100 & 8 & 1 & 0 & -1 & 3 & 3 & 2 & 11 & 5 & 18 & -10 & -4 & -16 & 11 & 11 & \text{ggF} \\
 8 & 100 & 50 & 36 & -10 & -27 & -2 & -3 & 14 & 0 & 24 & -10 & -4 & -19 & 17 & 17 & \text{VBF} \\
 1 & 50 & 100 & 65 & -2 & -13 & -17 & -19 & 8 & -14 & 16 & 4 & 0 & -6 & 19 & 19 & \text{ZH} \\
 0 & 36 & 65 & 100 & 0 & -4 & -16 & -18 & 7 & -13 & 14 & 4 & 0 & -5 & 18 & 18 & \text{WH} \\
 -1 & -10 & -2 & 0 & 100 & 10 & -4 & -4 & -1 & -4 & -1 & 3 & 1 & 2 & 1 & 1 & \text{ttH} \\
 3 & -27 & -13 & -4 & 10 & 100 & -7 & -7 & -4 & -7 & 9 & 3 & -5 & -16 & -2 & -2 & \text{bbH} \\
 3 & -2 & -17 & -16 & -4 & -7 & 100 & 100 & 56 & 99 & 36 & -86 & -15 & -29 & 10 & 10 & WW \\
 2 & -3 & -19 & -18 & -4 & -7 & 100 & 100 & 49 & 98 & 31 & -83 & -15 & -29 & 2 & 2 & ZZ \\
 11 & 14 & 8 & 7 & -1 & -4 & 56 & 49 & 100 & 67 & 77 & -82 & -11 & -15 & 88 & 88 & \gamma \gamma  \\
 5 & 0 & -14 & -13 & -4 & -7 & 99 & 98 & 67 & 100 & 46 & -91 & -16 & -28 & 24 & 24 & \text{ $Z\gamma$} \\
 18 & 24 & 16 & 14 & -1 & 9 & 36 & 31 & 77 & 46 & 100 & -73 & -20 & -70 & 70 & 70 & gg \\
 -10 & -10 & 4 & 4 & 3 & 3 & -86 & -83 & -82 & -91 & -73 & 100 & -1 & 46 & -48 & -48 & b\bar{b} \\
 -4 & -4 & 0 & 0 & 1 & -5 & -15 & -15 & -11 & -16 & -20 & -1 & 100 & 25 & -4 & -4 & c\bar{c} \\
 -16 & -19 & -6 & -5 & 2 & -16 & -29 & -29 & -15 & -28 & -70 & 46 & 25 & 100 & 1 & 1 & s\bar{s} \\
 11 & 17 & 19 & 18 & 1 & -2 & 10 & 2 & 88 & 24 & 70 & -48 & -4 & 1 & 100 & 100 & \tau \bar{\tau } \\
 11 & 17 & 19 & 18 & 1 & -2 & 10 & 2 & 88 & 24 & 70 & -48 & -4 & 1 & 100 & 100 & \mu \bar{\mu } \\
 \end{block}
\end{blockarray}
\nonumber
\ee
\caption{Correlation matrix between the expected Higgs  production and decay rates at  $\sqrt{s}=8$ TeV, expressed in percent, assuming the scale correlations $i)$. \label{tab:rho8} }
\end{table}

\begin{table}
\be
\small
\begin{blockarray}{*{17}{c<{\hspace{0pt}}}}
 \text{ggH} & \text{VBF} & \text{ZH} & \text{WH} & \text{ttH} & \text{bbH} & \,\,WW & ZZ & \gamma \gamma  &  Z \gamma &  gg & b\bar{b} & c\bar{c} & s\bar{s} & \tau \bar{\tau } & \mu \bar{\mu } &  \\
 \begin{block}{(*{15}{c<{\hspace{-10pt} }}c)c}
 100 & 9 & 5 & 4 & -2 & 3 & 3 & 2 & 10 & 5 & 18 & -10 & -4 & -15 & 10 & 10 & \text{ggF} \\
 9 & 100 & 50 & 38 & -12 & -24 & 1 & 0 & 16 & 4 & 28 & -14 & -6 & -23 & 18 & 18 & \text{VBF} \\
 5 & 50 & 100 & 56 & -6 & -11 & -13 & -15 & 14 & -9 & 26 & -4 & -3 & -16 & 23 & 23 & \text{ZH} \\
 4 & 38 & 56 & 100 & -4 & -3 & -13 & -15 & 13 & -10 & 24 & -3 & -3 & -14 & 22 & 22 & \text{WH} \\
 -2 & -12 & -6 & -4 & 100 & 9 & -4 & -4 & -2 & -4 & -2 & 4 & 1 & 3 & 0 & 0 & \text{ttH} \\
 3 & -24 & -11 & -3 & 9 & 100 & -8 & -8 & -6 & -8 & 8 & 4 & -5 & -16 & -3 & -3 & \text{bbH} \\
 3 & 1 & -13 & -13 & -4 & -8 & 100 & 100 & 56 & 99 & 36 & -86 & -15 & -29 & 10 & 10 & WW \\
 2 & 0 & -15 & -15 & -4 & -8 & 100 & 100 & 49 & 98 & 31 & -83 & -15 & -29 & 2 & 2 & ZZ \\
 10 & 16 & 14 & 13 & -2 & -6 & 56 & 49 & 100 & 67 & 77 & -82 & -11 & -15 & 88 & 88 & \gamma \gamma  \\
 5 & 4 & -9 & -10 & -4 & -8 & 99 & 98 & 67 & 100 & 46 & -91 & -16 & -28 & 24 & 24 & \text{ $Z\gamma$} \\
 18 & 28 & 26 & 24 & -2 & 8 & 36 & 31 & 77 & 46 & 100 & -73 & -20 & -70 & 70 & 70 & gg \\
 -10 & -14 & -4 & -3 & 4 & 4 & -86 & -83 & -82 & -91 & -73 & 100 & -1 & 46 & -48 & -48 & b\bar{b} \\
 -4 & -6 & -3 & -3 & 1 & -5 & -15 & -15 & -11 & -16 & -20 & -1 & 100 & 25 & -4 & -4 & c\bar{c} \\
 -15 & -23 & -16 & -14 & 3 & -16 & -29 & -29 & -15 & -28 & -70 & 46 & 25 & 100 & 1 & 1 & s\bar{s} \\
 10 & 18 & 23 & 22 & 0 & -3 & 10 & 2 & 88 & 24 & 70 & -48 & -4 & 1 & 100 & 100 & \tau \bar{\tau } \\
 10 & 18 & 23 & 22 & 0 & -3 & 10 & 2 & 88 & 24 & 70 & -48 & -4 & 1 & 100 & 100 & \mu \bar{\mu } \\
 \end{block}
\end{blockarray}
\nonumber
\ee
\caption{Correlation matrix between the expected Higgs  production and decay rates at  $\sqrt{s}=13$ TeV, expressed in percent, assuming the scale correlations $i)$. \label{tab:rho13} }
\end{table}

\begin{table}[h]
\be
\small
\begin{blockarray}{*{17}{c<{\hspace{0pt}}}}
 \text{ggH} & \text{VBF} & \text{ZH} & \text{WH} & \text{ttH} & \text{bbH} & \,\,WW & ZZ & \gamma \gamma  &  Z \gamma &  gg & b\bar{b} & c\bar{c} & s\bar{s} & \tau \bar{\tau } & \mu \bar{\mu } &  \\
 \begin{block}{(*{15}{c<{\hspace{-10pt} }}c)c}
 100 & 10 & 6 & 5 & -2 & 3 & 3 & 2 & 10 & 5 & 18 & -10 & -4 & -15 & 10 & 10 & \text{ggF} \\
 10 & 100 & 50 & 39 & -11 & -23 & 1 & -1 & 17 & 3 & 30 & -14 & -6 & -24 & 19 & 19 & \text{VBF} \\
 6 & 50 & 100 & 54 & -6 & -11 & -12 & -14 & 14 & -9 & 27 & -5 & -3 & -17 & 24 & 24 & \text{ZH} \\
 5 & 39 & 54 & 100 & -4 & -3 & -13 & -14 & 13 & -9 & 25 & -3 & -3 & -15 & 22 & 22 & \text{WH} \\
 -2 & -11 & -6 & -4 & 100 & 8 & -4 & -4 & -2 & -4 & -3 & 4 & 1 & 4 & 0 & 0 & \text{ttH} \\
 3 & -23 & -11 & -3 & 8 & 100 & -8 & -8 & -6 & -8 & 8 & 4 & -5 & -16 & -3 & -3 & \text{bbH} \\
 3 & 1 & -12 & -13 & -4 & -8 & 100 & 100 & 56 & 99 & 36 & -86 & -15 & -29 & 10 & 10 & WW \\
 2 & -1 & -14 & -14 & -4 & -8 & 100 & 100 & 49 & 98 & 31 & -83 & -15 & -29 & 2 & 2 & ZZ \\
 10 & 17 & 14 & 13 & -2 & -6 & 56 & 49 & 100 & 67 & 77 & -82 & -11 & -15 & 88 & 88 & \gamma \gamma  \\
 5 & 3 & -9 & -9 & -4 & -8 & 99 & 98 & 67 & 100 & 46 & -91 & -16 & -28 & 24 & 24 & \text{ $Z\gamma$} \\
 18 & 30 & 27 & 25 & -3 & 8 & 36 & 31 & 77 & 46 & 100 & -73 & -20 & -70 & 70 & 70 & gg \\
 -10 & -14 & -5 & -3 & 4 & 4 & -86 & -83 & -82 & -91 & -73 & 100 & -1 & 46 & -48 & -48 & b\bar{b} \\
 -4 & -6 & -3 & -3 & 1 & -5 & -15 & -15 & -11 & -16 & -20 & -1 & 100 & 25 & -4 & -4 & c\bar{c} \\
 -15 & -24 & -17 & -15 & 4 & -16 & -29 & -29 & -15 & -28 & -70 & 46 & 25 & 100 & 1 & 1 & s\bar{s} \\
 10 & 19 & 24 & 22 & 0 & -3 & 10 & 2 & 88 & 24 & 70 & -48 & -4 & 1 & 100 & 100 & \tau \bar{\tau } \\
 10 & 19 & 24 & 22 & 0 & -3 & 10 & 2 & 88 & 24 & 70 & -48 & -4 & 1 & 100 & 100 & \mu \bar{\mu } \\
 \end{block}
\end{blockarray}
\nonumber
\ee
\caption{Correlation matrix between the expected Higgs  production and decay rates at  $\sqrt{s}=14$ TeV, expressed in percent, assuming the scale correlations $i)$. \label{tab:rho14} }
\end{table}

\begin{table}[h]
\be
\small
\begin{blockarray}{*{17}{c<{\hspace{0pt}}}}
 \text{ggH} & \text{VBF} & \text{ZH} & \text{WH} & \text{ttH} & \text{bbH} & \,\,WW & ZZ & \gamma \gamma  &  Z \gamma &  gg & b\bar{b} & c\bar{c} & s\bar{s} & \tau \bar{\tau } & \mu \bar{\mu } &  \\
 \begin{block}{(*{15}{c<{\hspace{-10pt} }}c)c}
 100 & 7 & 30 & 26 & 82 & -36 & 3 & 2 & 11 & 5 & 19 & -11 & -4 & -16 & 11 & 11 & \text{ggF} \\
 7 & 100 & 55 & 40 & -7 & -28 & -3 & -4 & 13 & 0 & 23 & -9 & -4 & -18 & 16 & 16 & \text{VBF} \\
 30 & 55 & 100 & 97 & 34 & -32 & -20 & -22 & 6 & -17 & 14 & 6 & 0 & -4 & 19 & 19 & \text{ZH} \\
 26 & 40 & 97 & 100 & 32 & -19 & -19 & -21 & 6 & -16 & 13 & 6 & 0 & -3 & 18 & 18 & \text{WH} \\
 82 & -7 & 34 & 32 & 100 & -37 & -3 & -3 & 0 & -3 & 0 & 2 & 0 & 1 & 1 & 1 & \text{ttH} \\
 -36 & -28 & -32 & -19 & -37 & 100 & -8 & -8 & -4 & -8 & 11 & 2 & -6 & -19 & -1 & -1 & \text{bbH} \\
 3 & -3 & -20 & -19 & -3 & -8 & 100 & 100 & 56 & 99 & 36 & -86 & -15 & -29 & 10 & 10 & WW \\
 2 & -4 & -22 & -21 & -3 & -8 & 100 & 100 & 49 & 98 & 31 & -83 & -15 & -29 & 2 & 2 & ZZ \\
 11 & 13 & 6 & 6 & 0 & -4 & 56 & 49 & 100 & 67 & 77 & -82 & -11 & -15 & 88 & 88 & \gamma \gamma  \\
 5 & 0 & -17 & -16 & -3 & -8 & 99 & 98 & 67 & 100 & 46 & -91 & -16 & -28 & 24 & 24 & \text{ $Z\gamma$} \\
 19 & 23 & 14 & 13 & 0 & 11 & 36 & 31 & 77 & 46 & 100 & -73 & -20 & -70 & 70 & 70 & gg \\
 -11 & -9 & 6 & 6 & 2 & 2 & -86 & -83 & -82 & -91 & -73 & 100 & -1 & 46 & -48 & -48 & b\bar{b} \\
 -4 & -4 & 0 & 0 & 0 & -6 & -15 & -15 & -11 & -16 & -20 & -1 & 100 & 25 & -4 & -4 & c\bar{c} \\
 -16 & -18 & -4 & -3 & 1 & -19 & -29 & -29 & -15 & -28 & -70 & 46 & 25 & 100 & 1 & 1 & s\bar{s} \\
 11 & 16 & 19 & 18 & 1 & -1 & 10 & 2 & 88 & 24 & 70 & -48 & -4 & 1 & 100 & 100 & \tau \bar{\tau } \\
 11 & 16 & 19 & 18 & 1 & -1 & 10 & 2 & 88 & 24 & 70 & -48 & -4 & 1 & 100 & 100 & \mu \bar{\mu } \\
 \end{block}
\end{blockarray}
\nonumber
\ee
\caption{Correlation matrix between the expected Higgs  production and decay rates at  $\sqrt{s}=7$ TeV, expressed in percent, assuming the scale correlations $ii)$. \label{tab:rho7ii} }
\end{table}

\begin{table}[h]
\be
\small
\begin{blockarray}{*{17}{c<{\hspace{0pt}}}}
 \text{ggH} & \text{VBF} & \text{ZH} & \text{WH} & \text{ttH} & \text{bbH} & \,\,WW & ZZ & \gamma \gamma  &  Z \gamma &  gg & b\bar{b} & c\bar{c} & s\bar{s} & \tau \bar{\tau } & \mu \bar{\mu } &  \\
 \begin{block}{(*{15}{c<{\hspace{-10pt} }}c)c}
 100 & 1 & 25 & 22 & 82 & -34 & 3 & 2 & 11 & 5 & 19 & -11 & -4 & -16 & 11 & 11 & \text{ggF} \\
 1 & 100 & 56 & 40 & -16 & -27 & -2 & -3 & 14 & 0 & 24 & -10 & -4 & -19 & 17 & 17 & \text{VBF} \\
 25 & 56 & 100 & 97 & 26 & -30 & -20 & -22 & 9 & -16 & 18 & 4 & -1 & -7 & 22 & 22 & \text{ZH} \\
 22 & 40 & 97 & 100 & 25 & -16 & -19 & -21 & 8 & -16 & 17 & 4 & 0 & -6 & 20 & 20 & \text{WH} \\
 82 & -16 & 26 & 25 & 100 & -35 & -3 & -3 & 0 & -3 & -1 & 2 & 1 & 1 & 1 & 1 & \text{ttH} \\
 -34 & -27 & -30 & -16 & -35 & 100 & -8 & -8 & -5 & -8 & 11 & 3 & -6 & -19 & -2 & -2 & \text{bbH} \\
 3 & -2 & -20 & -19 & -3 & -8 & 100 & 100 & 56 & 99 & 36 & -86 & -15 & -29 & 10 & 10 & WW \\
 2 & -3 & -22 & -21 & -3 & -8 & 100 & 100 & 49 & 98 & 31 & -83 & -15 & -29 & 2 & 2 & ZZ \\
 11 & 14 & 9 & 8 & 0 & -5 & 56 & 49 & 100 & 67 & 77 & -82 & -11 & -15 & 88 & 88 & \gamma \gamma  \\
 5 & 0 & -16 & -16 & -3 & -8 & 99 & 98 & 67 & 100 & 46 & -91 & -16 & -28 & 24 & 24 & \text{ $Z\gamma$} \\
 19 & 24 & 18 & 17 & -1 & 11 & 36 & 31 & 77 & 46 & 100 & -73 & -20 & -70 & 70 & 70 & gg \\
 -11 & -10 & 4 & 4 & 2 & 3 & -86 & -83 & -82 & -91 & -73 & 100 & -1 & 46 & -48 & -48 & b\bar{b} \\
 -4 & -4 & -1 & 0 & 1 & -6 & -15 & -15 & -11 & -16 & -20 & -1 & 100 & 25 & -4 & -4 & c\bar{c} \\
 -16 & -19 & -7 & -6 & 1 & -19 & -29 & -29 & -15 & -28 & -70 & 46 & 25 & 100 & 1 & 1 & s\bar{s} \\
 11 & 17 & 22 & 20 & 1 & -2 & 10 & 2 & 88 & 24 & 70 & -48 & -4 & 1 & 100 & 100 & \tau \bar{\tau } \\
 11 & 17 & 22 & 20 & 1 & -2 & 10 & 2 & 88 & 24 & 70 & -48 & -4 & 1 & 100 & 100 & \mu \bar{\mu } \\
 \end{block}
\end{blockarray}
\nonumber
\ee
\caption{Correlation matrix between the expected Higgs  production and decay rates at  $\sqrt{s}=8$ TeV, expressed in percent, assuming the scale correlations $ii)$. \label{tab:rho8ii} }
\end{table}

\begin{table}[h]
\be
\small
\begin{blockarray}{*{17}{c<{\hspace{0pt}}}}
 \text{ggH} & \text{VBF} & \text{ZH} & \text{WH} & \text{ttH} & \text{bbH} & \,\,WW & ZZ & \gamma \gamma  &  Z \gamma &  gg & b\bar{b} & c\bar{c} & s\bar{s} & \tau \bar{\tau } & \mu \bar{\mu } &  \\
 \begin{block}{(*{15}{c<{\hspace{-10pt} }}c)c}
 100 & -18 & 7 & 5 & 80 & -28 & 4 & 3 & 11 & 5 & 20 & -11 & -4 & -17 & 11 & 11 & \text{ggF} \\
 -18 & 100 & 65 & 49 & -41 & -17 & 1 & 0 & 16 & 4 & 28 & -14 & -6 & -23 & 17 & 17 & \text{VBF} \\
 7 & 65 & 100 & 96 & -7 & -18 & -17 & -20 & 18 & -12 & 34 & -5 & -4 & -21 & 31 & 31 & \text{ZH} \\
 5 & 49 & 96 & 100 & -4 & -5 & -17 & -20 & 16 & -13 & 31 & -3 & -3 & -19 & 29 & 29 & \text{WH} \\
 80 & -41 & -7 & -4 & 100 & -30 & -3 & -3 & -1 & -3 & -2 & 3 & 1 & 2 & 0 & 0 & \text{ttH} \\
 -28 & -17 & -18 & -5 & -30 & 100 & -10 & -10 & -7 & -10 & 9 & 5 & -6 & -19 & -4 & -4 & \text{bbH} \\
 4 & 1 & -17 & -17 & -3 & -10 & 100 & 100 & 56 & 99 & 36 & -86 & -15 & -29 & 10 & 10 & WW \\
 3 & 0 & -20 & -20 & -3 & -10 & 100 & 100 & 49 & 98 & 31 & -83 & -15 & -29 & 2 & 2 & ZZ \\
 11 & 16 & 18 & 16 & -1 & -7 & 56 & 49 & 100 & 67 & 77 & -82 & -11 & -15 & 88 & 88 & \gamma \gamma  \\
 5 & 4 & -12 & -13 & -3 & -10 & 99 & 98 & 67 & 100 & 46 & -91 & -16 & -28 & 24 & 24 & \text{ $Z\gamma$} \\
 20 & 28 & 34 & 31 & -2 & 9 & 36 & 31 & 77 & 46 & 100 & -73 & -20 & -70 & 70 & 70 & gg \\
 -11 & -14 & -5 & -3 & 3 & 5 & -86 & -83 & -82 & -91 & -73 & 100 & -1 & 46 & -48 & -48 & b\bar{b} \\
 -4 & -6 & -4 & -3 & 1 & -6 & -15 & -15 & -11 & -16 & -20 & -1 & 100 & 25 & -4 & -4 & c\bar{c} \\
 -17 & -23 & -21 & -19 & 2 & -19 & -29 & -29 & -15 & -28 & -70 & 46 & 25 & 100 & 1 & 1 & s\bar{s} \\
 11 & 17 & 31 & 29 & 0 & -4 & 10 & 2 & 88 & 24 & 70 & -48 & -4 & 1 & 100 & 100 & \tau \bar{\tau } \\
 11 & 17 & 31 & 29 & 0 & -4 & 10 & 2 & 88 & 24 & 70 & -48 & -4 & 1 & 100 & 100 & \mu \bar{\mu } \\
 \end{block}
\end{blockarray}
\nonumber
\ee
\caption{Correlation matrix between the expected Higgs  production and decay rates at  $\sqrt{s}=13$ TeV, expressed in percent, assuming the scale correlations $ii)$. \label{tab:rho13ii} }
\end{table}

\begin{table}[h]
\be
\small
\begin{blockarray}{*{17}{c<{\hspace{0pt}}}}
 \text{ggH} & \text{VBF} & \text{ZH} & \text{WH} & \text{ttH} & \text{bbH} & \,\,WW & ZZ & \gamma \gamma  &  Z \gamma &  gg & b\bar{b} & c\bar{c} & s\bar{s} & \tau \bar{\tau } & \mu \bar{\mu } &  \\
 \begin{block}{(*{15}{c<{\hspace{-10pt} }}c)c}
 100 & -19 & 5 & 3 & 80 & -27 & 4 & 3 & 12 & 5 & 20 & -12 & -4 & -17 & 11 & 11 & \text{ggF} \\
 -19 & 100 & 67 & 51 & -43 & -14 & 1 & -1 & 16 & 3 & 29 & -14 & -6 & -23 & 18 & 18 & \text{VBF} \\
 5 & 67 & 100 & 96 & -11 & -16 & -16 & -19 & 19 & -11 & 36 & -6 & -5 & -23 & 32 & 32 & \text{ZH} \\
 3 & 51 & 96 & 100 & -8 & -4 & -16 & -19 & 17 & -12 & 33 & -5 & -4 & -20 & 30 & 30 & \text{WH} \\
 80 & -43 & -11 & -8 & 100 & -29 & -3 & -3 & -1 & -3 & -2 & 3 & 1 & 3 & 0 & 0 & \text{ttH} \\
 -27 & -14 & -16 & -4 & -29 & 100 & -10 & -10 & -8 & -10 & 9 & 5 & -6 & -19 & -4 & -4 & \text{bbH} \\
 4 & 1 & -16 & -16 & -3 & -10 & 100 & 100 & 56 & 99 & 36 & -86 & -15 & -29 & 10 & 10 & WW \\
 3 & -1 & -19 & -19 & -3 & -10 & 100 & 100 & 49 & 98 & 31 & -83 & -15 & -29 & 2 & 2 & ZZ \\
 12 & 16 & 19 & 17 & -1 & -8 & 56 & 49 & 100 & 67 & 77 & -82 & -11 & -15 & 88 & 88 & \gamma \gamma  \\
 5 & 3 & -11 & -12 & -3 & -10 & 99 & 98 & 67 & 100 & 46 & -91 & -16 & -28 & 24 & 24 & \text{ $Z\gamma$} \\
 20 & 29 & 36 & 33 & -2 & 9 & 36 & 31 & 77 & 46 & 100 & -73 & -20 & -70 & 70 & 70 & gg \\
 -12 & -14 & -6 & -5 & 3 & 5 & -86 & -83 & -82 & -91 & -73 & 100 & -1 & 46 & -48 & -48 & b\bar{b} \\
 -4 & -6 & -5 & -4 & 1 & -6 & -15 & -15 & -11 & -16 & -20 & -1 & 100 & 25 & -4 & -4 & c\bar{c} \\
 -17 & -23 & -23 & -20 & 3 & -19 & -29 & -29 & -15 & -28 & -70 & 46 & 25 & 100 & 1 & 1 & s\bar{s} \\
 11 & 18 & 32 & 30 & 0 & -4 & 10 & 2 & 88 & 24 & 70 & -48 & -4 & 1 & 100 & 100 & \tau \bar{\tau } \\
 11 & 18 & 32 & 30 & 0 & -4 & 10 & 2 & 88 & 24 & 70 & -48 & -4 & 1 & 100 & 100 & \mu \bar{\mu } \\
 \end{block}
\end{blockarray}
\nonumber
\ee
\caption{Correlation matrix between the expected Higgs  production and decay rates at  $\sqrt{s}=14$ TeV, expressed in percent, assuming the scale correlations $ii)$. \label{tab:rho14ii} }
\end{table}

\newpage

{\renewcommand{\arraystretch}{1.5}


\end{document}